\documentclass[12pt,letterpaper]{article}
\usepackage[top=3cm, bottom=3cm, left=1.5cm, right=2cm]{geometry}
\usepackage[usenames,dvipsnames,svgnames]{xcolor}
\definecolor{darkblue}{rgb}{0.0,0.1,0.3} 
\definecolor{darkgreen}{rgb}{0,0.65,0}
\definecolor{dblue4}{rgb}{0.06,0.31,0.55} 
\definecolor{nicered}{rgb}{0.7,0.1,0.1}
\definecolor{nicegreen}{rgb}{0.1,0.5,0.1}

\usepackage[numbers,sort&compress]{natbib}
\usepackage[utf8x]{inputenc}
\usepackage{amsmath,amssymb,amsfonts,amsthm}
\usepackage{xcolor}
\usepackage[colorlinks=true,citecolor= nicegreen,linkcolor=nicered]{hyperref}

\usepackage{verbatim}
\usepackage{graphicx}
\graphicspath{{figures/}}
\usepackage{cancel}
\usepackage{tipa}
\usepackage{array}   
\newcolumntype{L}{>{$}l<{$}} 
\newcolumntype{R}{>{$}r<{$}} 
\newcolumntype{C}{>{$}c<{$}} 
\usepackage{booktabs}
\usepackage{tabularx}
\usepackage{multirow}
\usepackage{dsfont}
\newcolumntype{Y}{>{\centering\arraybackslash}X}
\usepackage{tikz}

\def\c{,\allowbreak}

\title{Effective Dirac neutrino mass operator in the Standard Model  with a local Abelian extension}

\author{ Diego Restrepo\footnote{\href{mailto:restrepo@udea.edu.co}{restrepo@udea.edu.co}}\ \ and David Suarez\footnote{\href{mailto:david.suarezr@udea.edu.co}{david.suarezr@udea.edu.co}}\\
\textit{\small Instituto de Física, Universidad de Antioquia}\\
\textit{\small  Calle 70 \# 52-21, Apartado Aéreo 1226, Medellín, Colombia}
}
\date{}
\begin{document}
\maketitle

\begin{abstract}
We present 48 types of solutions to the anomaly cancellation conditions of local Abelian extensions of the Standard Model (SM) with right-handed singlet chiral fermions. At least two of them acquire effective light Dirac neutrino masses, while the others get heavy masses from the spontaneous symmetry breaking of the local Abelian symmetry, forming a dark sector with multi-component and multi-generational fermionic dark matter. The corresponding effective Dirac neutrino mass operator can be realized at tree-level or radiatively  by introducing extra scalars, and in some cases after imposing extra scotogenic conditions. The Dirac Zee model with Dirac fermionic dark matter is presented as an example of model where the neutrino and dark matter phenomenology are basically independent of each other.
\end{abstract}

\section{Introduction}
We can relate fermionic dark matter (DM) with neutrinos in two ways, through the fermions inside the loops in scotogenic models~\cite{Ma:2006km} or through anomaly-free gauge extensions of the SM with the extra chiral fermions required to cancel out the anomalies~\cite{Patra:2016ofq}. With Dirac neutrinos, we can have the two ways simultaneously. In fact, regarding neutrinos, the oscillation data is compatible with
both Majorana or Dirac neutrino masses. To explain Dirac neutrino masses, right-handed neutrinos have to be introduced. Additionally, an extra local symmetry is also required to guarantee proper total lepton number conservation. Moreover, to have a consistent anomaly-free local symmetry, extra right-handed singlet chiral fields are required in a dark sector which may include dark matter candidates.  

In this work we propose a general framework for the systematic study of Dirac neutrino masses in the context of a gauge Abelian symmetry with multi-component dark matter and massive gauge bosons, including the possibility of dark photons. 

By using an effective operator approach with until 10  extra singlet chiral fields, we found a well defined set of 48 types of solutions with effective Dirac neutrino masses, multi-component and multi-generational dark matter, and either a dark photon or a heavy gauge boson mediator. 

This work is organized as follows.
The anomaly conditions for the gauge Abelian symmetry are presented in section \ref{sec:anomaly}, and expressed in terms of the charges of a set of right-handed singlet chiral fields.
In section~\ref{sec:Dirac}, a subset of them are used as the right-handed neutrino  component of the effective Dirac neutrino mass operator. Then, the charge of the complex singlet scalar that breaks spontaneously the gauge Abelian symmetry is fixed. In section~\ref{sec:darkmatter} we search for the solutions in which the remaining subset of right-handed singlet chiral fermions obtain heavy masses from the singlet scalar vacuum expectation value. There, they are classified in 48 types of solutions and their main features are highlighted. In section~\ref{sec:zee} the solution with the least number of chiral fields is discussed and used as an example of the realization of the effective Dirac neutrino mass operator through SM scalar extensions.
Finally, in section~\ref{sec:con} our conclusions are presented.

\section{Anomaly conditions} 
\label{sec:anomaly}
We consider an extension of the SM with an additional Abelian gauge symmetry and extra right-handed chiral fields singlets under the SM $\operatorname{SU}(3)_c\otimes \operatorname{SU}(2)_L\otimes \operatorname{U}(1)_Y$ group. We assume that they do not form vector-like pairs so that all of them are massless before the spontaneous symmetry breaking of the Abelian gauge symmetry. On the other hand, we consider a set of $N$ integers, $n_1,n_2,\ldots,n_N$, which satisfy the following Diophantine equations
\begin{align} \label{eq:NN3}
 \sum_{\rho=1}^{N}n_{\rho}=0 \qquad \text{and} \qquad \sum_{\rho=1}^{N}n_{\rho}^3=0\,.
\end{align}
In \cite{Bernal:2021ezl} it is shown that
\begin{itemize}
    \item The extra gauge Abelian symmetry can be identified as one \emph{dark} symmetry, $\operatorname{U}(1)_D$, with $N_{\text{chiral}}=N$ right-handed singlet chiral fermions with $D$-charges $n_1,n_2,\ldots n_N$.
    \item If the set of integers has one integer, $m$, repeated three times, the extra gauge Abelian symmetry can be identified as one \emph{active} symmetry, $\operatorname{U}(1)_X$, with $N_{\text{chiral}}=N-3$ right-handed singlet chiral fermions with $X$-charges $n_1,n_2,\ldots n_{N_{\text{chiral}}}$.  The SM right-handed chiral fermions $X$-charges are denoted with the same name of the field%
\footnote{$Q$ and $L$ are the $X$-charges of the fermion doublets $Q^{\dagger}$ and $L^{\dagger}$, respectively.}. They can be written in terms of $m$ and a free parameter that we choose to be the $X$-charge of the conjugate of the SM lepton doublet, $L$, such that
\begin{align}
  u=&\frac{4 L}{3}-m\,,& d=m-\frac{2 L}{3}& \,,& Q=& -\frac{L}{3}\,,& e=& m-2 L\,,&
    h =& L-m\,,
\end{align}
where $h$ is  the SM Higgs  $X$-charge, and $Q$ is the $X$-charge of the conjugate of the SM quark doublet.
\end{itemize}

We are interested in \emph{chiral} solutions: the ones in which there is no $D$ or $X$ opposite sign charges for the right-handed chiral fermions. This is motivated by the fact that there are only chiral fermions in the SM.

The simple solution to obtain an active symmetry,  $\operatorname{U}(1)_X$, is just to have $m=1$ and three right-handed chiral fermions. When $h=0$, such that $L=1$, this is the well known $\operatorname{U}(1)_{B-L}$ set of $N$ integers satisfying eq.~\eqref{eq:NN3}: $(-1\c-1\c-1\c1\c1\c1)$, and any multiplicative factor of it: $(-m\c-m\c-m\c m\c m\c m)$\footnote{This solutions cannot be interpreted as a dark symmetry because it would contain three pairs of vector-like singlet fermions.}. Note that we can make the direct sum of this simple solution with any other solution with $N_{\text{chiral}}$ integers that does not contains $m$, to obtain an active gauge Abelian symmetry with three extra right handed chiral fermions with $X$-charges $-m$~\cite{Bernal:2021ppq}. The same procedure can be used to obtain active symmetries for any chiral solution of eq.~\eqref{eq:NN3}, by identifying a subset of its integers as $m$. For example, from the well known solution $(1,1,1,-4,-4,5)$~\cite{Appelquist:2002mw,Montero:2007cd} of the eq.~\eqref{eq:NN3}, we can build the following extra active solutions with their corresponding $m$
\begin{align}
\label{eq:dirsum}
    (1,1,1,-4,-4,5)\oplus (-m,-m,-m, m,m,m)=&(1,1,1,-4,-4,5,-m,-m,-m, m,m,m),\text{ with $m\ne \pm 1$}\nonumber\\
    (1,1,1,-4,-4,5)\oplus (5,5,-5, -5)=&(1,1,1,-4,-4,5,5,5,-5,5),\text{ with $m=5$}\nonumber\\
    (1,1,1,-4,-4,5)\oplus (-4,4)=&(1,1,1,-4,-4,-4,4,5),\text{ with $m=-4$}\,.
\end{align}
Finally, for any given solution, we also have a new solution from the direct sum with itself. For example
\begin{align}
\label{eq:double}
        (1,1,1,-4,-4,5)\oplus (1,1,1,-4,-4,5)=&(1,1,1,1,1,1,-4,-4,-4,-4,5,5)
        ,\text{ with $m=1$ or $m=-4$}.
\end{align}

In the next section we will apply phenomenological conditions upon all  the solutions with $5\le N\le 12$ integers  obtained with the method explained in~\cite{Costa:2019zzy} and implemented in the python package \texttt{anomalies}~\cite{diego_restrepo_2021_5526558}\footnote{https://pypi.org/project/anomalies/}. The corresponding dataset with $390\,074$ solutions to eqs.~\eqref{eq:NN3}, with a maximum absolute value of 30, is in~\cite{diego_restrepo_2021_5526707}. They are complemented with the direct sums as in eqs.~\eqref{eq:dirsum} and \eqref{eq:double} for each solution while keeping the maximum number of integers until 12.

\section{Effective Dirac neutrino mass operator} 
\label{sec:Dirac}
In this section, we look for anomaly-free Abelian gauge extensions of the SM, with  right-handed singlet chiral fermions, realizing the effective Dirac neutrino mass operator at a given dimension~\cite{Cleaver:1997nj, Gu:2006dc}.
In the two-component spinor notation, this can be written as
\begin{align} 
\label{eq:nmo56}
    \mathcal{L}_{\text{eff}} = h_{\nu}^{\alpha i} \, \left( \nu_{R\alpha}\right)^{\dagger} \, \epsilon_{ab} \, L_i^a \, H^b \left(\frac{S^*}{\Lambda}\right)^\delta + \text{H.c.},\qquad \text{with $i=1,2,3$}\,,
\end{align}
and $\delta = 1,2,\ldots$ for dimension $4+\delta$ operators.
Here $h_{\nu}^{\alpha i}$ are dimensionless induced couplings with at least a rank-2 matrix structure, $\nu_{R\alpha}$ are at least two right-handed neutrinos ($\alpha=1,2,\ldots$) with equal Abelian charge $\nu$ extracted from the full set of right-handed singlet chiral fields, $L_{i}$ are the lepton doublets with Abelian charge $-L$, $H$ is the SM Higgs doublet with Abelian charge $h=L-m$, $S$ is the complex singlet scalar responsible for the spontaneous symmetry breaking (SSB) of the anomaly-free gauge symmetry with Abelian charge
\begin{align}
   \label{eq:effcon}
    s=-(\nu+m)/\delta\,.
\end{align}
Finally, $\Lambda$ is a scale of new physics.

In general, after the SSB, several discrete symmetry are left, which may guarantee the stability of at least one potential DM candidate~\cite{Batell:2010bp}. $m\ne 0$ define an active symmetry where the collider restrictions on $Z'$ are rather strong because its allowed couplings to both light quarks and light leptons, while $m=0$ corresponds to a dark symmetry with the corresponding dark photon.

The tree-level realization of the renormalizable operator for Dirac neutrino masses through the Higgs mechanism is obtained when $\delta=0$, such that $\nu=-m$. Therefore, it is only possible for active symmetries. The simple realization is for the $\operatorname{U}(1)_{B-L}$ solution $(-1,-1,-1,1,1,1)$ with $m=1$, $\nu=-1$, and $s=3$ to avoid a Majorana mass for $\nu_R$. Therefore, the remnant $Z_3$ symmetry guarantees the Diracness of the tree-level neutrino masses. However, such a solution imply very tiny Yukawa coupling $h_\nu^{\alpha i}\sim\mathcal{O}(10^{-11})$, and therefore we do not consider this kind of solutions here because their lack of explanation for the smallness of neutrino masses.

For $\delta>0$ we can have tree-level~\cite{Bernal:2021ppq} or radiative~\cite{Bernal:2021ezl} realizations of the non-renormalizable effective Dirac neutrino mass operator. The specific realization can imply additional phenomenological conditions besides eq.~\eqref{eq:effcon}~\cite{Bernal:2021ezl}. However, there are realizations that only involve extra scalars beyond the SM~\cite{Saad:2019bqf} which do not affect the anomaly cancellation conditions. Therefore, any found solution satisfying eq.~\eqref{eq:effcon} can be embedded into a SM scalar extension which realize the effective Dirac neutrino mass operator, like the tree-level Type-II Dirac seesaw~\cite{Bernal:2021ppq}. In this way, we can focus into find the  self-consistent solutions with at least two repeated charges, $\nu$, which are compatible with the effective Dirac neutrino mass operator~\eqref{eq:nmo56}. This self-consistent solutions are defined  such that the only particles beyond the SM are $\nu_{R\alpha}$, $S$ and the set of  singlet chiral fermions which acquires masses from $\langle S\rangle$ and that can play the role of (multi-component) dark matter. The solutions until $N=9$ were already found in~\cite{Bernal:2021ppq} in the context of type-II Dirac seesaw realizations of the effective Dirac neutrino mass operator for $\delta=1$. Here, we extend the search of self-consistent solutions in the more general framework of the effective operator  until $N=12$ and all the relevant values for $\delta$.

\section{Dark sector with multi-component and multi-generational dark matter}
\label{sec:darkmatter}
Starting from the extended dataset for the solutions with $N$ integers to the Diophantine equations~\eqref{eq:NN3}, we apply the following steps
\begin{itemize}
    \item Check that the solution has two (three) repeated integers to be identified as $\nu$ and fix $N_\nu=2$  ($N_\nu=3$).
    \item For $\delta=1,2,\ldots$ and all the possible combinations for $m$ and $\nu$ in the solution, including $m=0$, find the $s$ value compatible with the effective Dirac neutrino mass operator of D-$(4+\delta)$ according to eq.~\eqref{eq:effcon}.
    \item Interpret the integers in the solution that are different from $m$ and $\nu$ as the $D$-charges for $m=0$ or the $X$-charges for $m\ne 0$, of a set of singlet chiral fermions: $\psi_i$, $i=1,\ldots, N_{\text{chiral}}-N_{\nu}$.  Then select the solutions for which the condition 
    \begin{align}
        |n_i+n_j|=|s|,
    \end{align}
    which guarantees that all the singlet chiral fermions, $\psi_i$, acquire masses after the spontaneous symmetry breaking of the gauge Abelian symmetry through $\langle S\rangle$.
\end{itemize}

\begin{figure}
    \centering
    \includegraphics[scale=0.7]{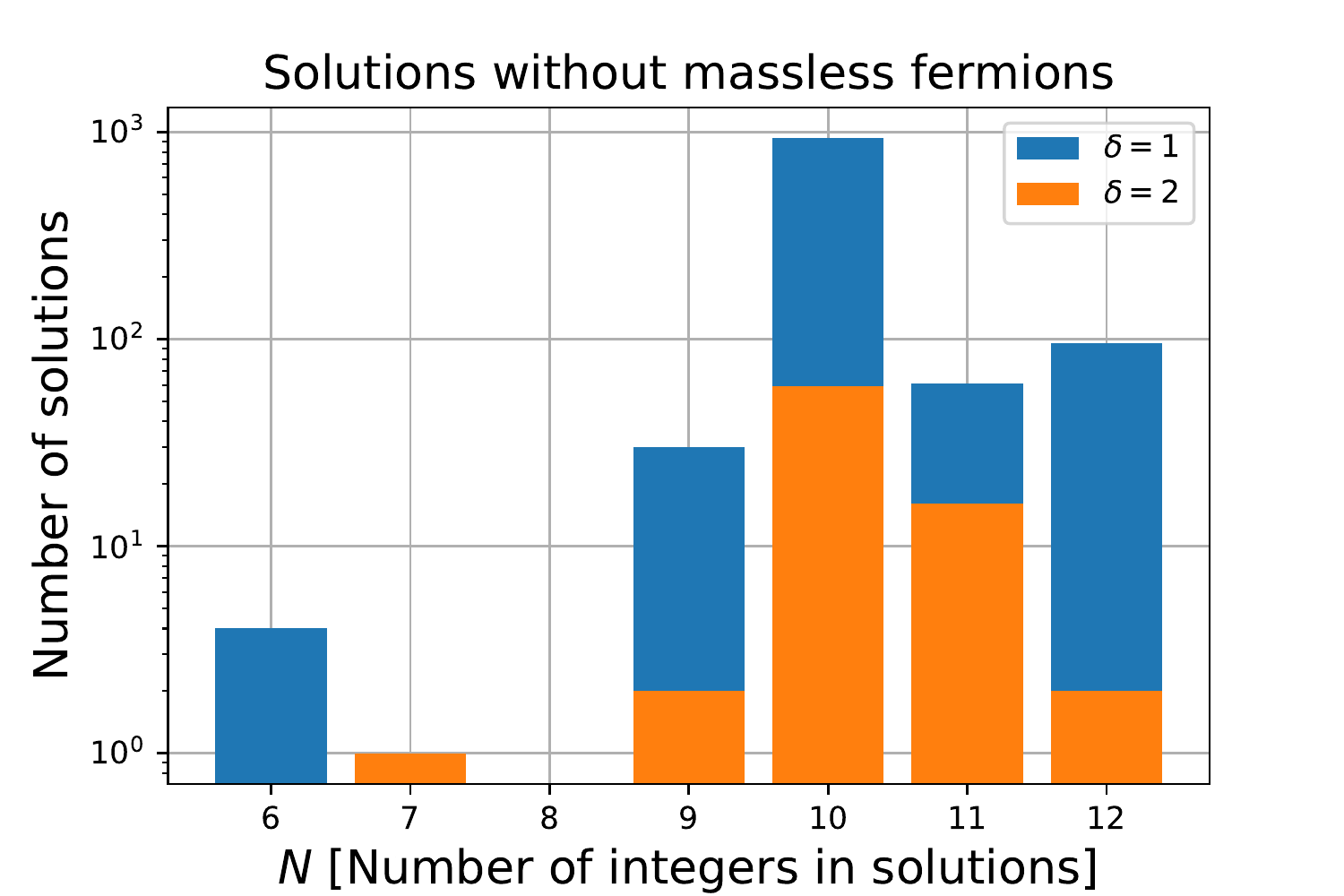}
     \caption{Distribution of solutions with $N$ integers to the Diophantine equations~\eqref{eq:NN3} which allow the effective Dirac neutrino mass operator at D-$(4+\delta)$ for at least two right-handed neutrinos and have non-vanishing Dirac o Majorana masses for the other singlet chiral fermions in the solution.}
    \label{fig:solutions}
\end{figure}

We found $1\,122$ solutions where all the singlet chiral fermions have either one effective mass through the operator in eq.~\eqref{eq:nmo56} or from the vacuum expectation value of $S$ after the spontaneous symmetry breaking of either the dark or the active Abelian gauge symmetry. As shown in Fig.~\ref{fig:solutions}, most of the solutions are for $\delta=1$ and $N=10$. 

Concerning the solutions with multi-component DM, we also explore the cases which feature at least two DM candidates with \emph{unconditional} stability~\cite{Yaguna:2019cvp,Bernal:2021ppq}. This happens when the remnant symmetry $\mathbb{Z}_{|s|}$ contains two subgroups $\mathbb{Z}_p\otimes\mathbb{Z}_q$ with  $p$ and $q$ coprimes and $|s|=pq$, which guarantee the stability of each lightest state under $\mathbb{Z}_p$ and $\mathbb{Z}_q$ respectively. For the two DM candidates associated to the set of chiral fields $\psi_i$ and $\chi_j$, we consider below the following two possibilities for $|s|$~\cite{Yaguna:2019cvp,Bernal:2021ppq}
\begin{itemize}
    \item $\mathbb{Z}_{6}\cong \mathbb{Z}_2\otimes\mathbb{Z}_3$: solutions with at least a set of chiral fields with $\psi_i \sim\left[\omega_{6}^{2} \vee \omega_{6}^{4}\right]$ under $\mathbb{Z}_{6}$, and at least a set of chiral fields with $\chi_i \sim \omega_{6}^3$ under $\mathbb{Z}_{6}$,
    \item $\mathbb{Z}_{14}\cong \mathbb{Z}_2\otimes\mathbb{Z}_7$: solutions with at least a set of chiral fields with $\psi_i \sim\left[\omega_{14}^{2}  \vee \omega_{14}^{6} \vee \omega_{14}^{8} \vee \omega_{14}^{10} \vee \omega_{14}^{12}\right]$ under $\mathbb{Z}_{14}$ and at least a set of chiral fields with $\chi_i \sim \omega_{14}^7$ under $\mathbb{Z}_{14}$,
\end{itemize}
where $\omega_{|s|}=\operatorname{e}^{i\, 2\pi/|s|}$.

The solutions can be classified in 48 types depending on the dimension of the effective Dirac neutrino mass operator, D-$(4+\delta)$, for $\delta=1,2$ values\footnote{For larger $\delta$ there a few other solutions with similar features}; the number of Dirac or Majorana massive singlet fermions; the multiplicity of generations for each one of them; and the unconditional stability of at least two of the multi-component dark matter candidates. The most representative  solutions (with the lower possible maximum charge) for each type, are presented in Table~\ref{tab:sltns}.

\begin{table}
  \centering
  \begin{scriptsize}
  \begin{tabular}{L|RC|RRR|R|RRRR}
\toprule
                                           \text{Solution} & N &N_{\text{chiral}} &   m & \nu & \delta &                s  & N_D & N_M & G_D & G_M \\
\midrule
                              (1, -2, -3, 5, 5, -6) &  6 &  6 &   0 &   5 &      1 &               -5 &   2 &   0 &   1 &   0 \\
                           (3, 3, 3, -5, -5, -7, 8) &  7 &  4 &   3 &  -5 &      2 &                1 &   1 &   0 &   1 &   0 \\
                    (1, -2, 3, 4, 6, -7, -7, -7, 9) &  9 &  9 &   0 &  -7 &      1 &                7 &   3 &   0 &   1 &   0 \\
                  (1, 1, -4, -5, 9, 9, 9, -10, -10) &  9 &  9 &   0 &   9 &      1 &               -9 &   3 &   0 &   2 &   0 \\
                   (1, 2, -6, -6, -6, 8, 9, 9, -11) &  9 &  6 &  -6 &   9 &      1 &               -3 &   2 &   0 &   1 &   0 \\
                (1, -3, 8, 8, 8, -12, -12, -17, 19) &  9 &  6 &   8 & -12 &      2 &                2 &   2 &   1 &   1 &   1 \\
              (8, 8, 8, -12, -12, 15, -17, -23, 25) &  9 &  6 &   8 & -12 &      2 &                2 &   2 &   0 &   1 &   0 \\
                (1, -2, -2, 3, 3, -4, -4, 6, 6, -7) & 10 & 10 &   0 &   6 &      1 &               -6 &   3 &   2 &   2 &   2 \\
                (1, -2, -2, 3, 4, -5, -5, 7, 7, -8) & 10 & 10 &   0 &  -5 &      1 &   \boldsymbol{5} &   4 &   0 &   2 &   0 \\
                (1, -2, -2, 3, 5, -6, -6, 8, 8, -9) & 10 & 10 &   0 &  -6 &      1 &                6 &   4 &   0 &   2 &   0 \\
                 (2, 2, 3, 4, 4, -5, -6, -6, -7, 9) & 10 & 10 &   0 &   2 &      1 &  \boldsymbol{-2} &   4 &   2 &   2 &   2 \\
                (1, 1, 5, 5, 5, -6, -6, -6, -9, 10) & 10 & 10 &   0 &   1 &      1 &               -1 &   4 &   0 &   3 &   0 \\
               (2, 2, 4, 4, -7, -7, -9, -9, 10, 10) & 10 & 10 &   0 &  10 &      2 &               -5 &   3 &   0 &   2 &   0 \\
                (1, 2, 2, -3, 6, 6, -8, -8, -9, 11) & 10 & 10 &   0 &  -8 &      1 &                8 &   4 &   1 &   2 &   1 \\
             (1, -2, -3, 5, 6, -8, -9, 11, 11, -12) & 10 & 10 &   0 &  11 &      1 &              -11 &   4 &   0 &   1 &   0 \\
              (1, 1, -3, 4, 4, -7, 8, -10, -10, 12) & 10 & 10 &   0 & -10 &      2 &                5 &   4 &   0 &   2 &   0 \\
            (1, 1, -2, -2, -4, 6, -10, 11, 12, -13) & 10 & 10 &   0 &  -2 &      1 &                2 &   3 &   2 &   1 &   2 \\
               (3, 4, 4, 4, 4, -5, -8, -8, -11, 13) & 10 & 10 &   0 &  -8 &      1 &                8 &   2 &   4 &   1 &   4 \\
             (4, 4, 5, 6, 6, -9, -10, -10, -11, 15) & 10 & 10 &   0 &   6 &      1 &               -6 &   4 &   0 &   2 &   0 \\
           (1, -2, -4, 7, 7, -10, -12, 14, 14, -15) & 10 & 10 &   0 &  14 &      1 & \boldsymbol{-14} &   3 &   2 &   1 &   2 \\
             (1, 2, 2, -3, 4, -6, 12, -13, -14, 15) & 10 & 10 &   0 &   2 &      1 &  \boldsymbol{-2} &   4 &   1 &   1 &   1 \\
              (1, 4, 4, -7, 8, 8, -9, -12, -12, 15) & 10 & 10 &   0 &   8 &      1 &               -8 &   4 &   2 &   2 &   2 \\
            (1, 2, 2, -9, -9, 16, 16, 17, -18, -18) & 10 & 10 &   0 & -18 &      1 &               18 &   3 &   2 &   2 &   2 \\
          (1, -3, -6, 7, -10, 11, -16, 18, 18, -20) & 10 & 10 &   0 &  18 &      2 &               -9 &   4 &   0 &   1 &   0 \\
           (1, -4, 5, -6, -6, 10, -14, 15, 20, -21) & 10 & 10 &   0 &  -6 &      1 &                6 &   4 &   0 &   1 &   0 \\
          (2, -3, -6, 7, 12, -14, -14, 17, 20, -21) & 10 & 10 &   0 & -14 &      1 &  \boldsymbol{14} &   4 &   1 &   1 &   1 \\
             (3, 6, 6, -7, 8, 8, -14, -14, -17, 21) & 10 & 10 &   0 & -14 &      1 &  \boldsymbol{14} &   4 &   1 &   2 &   1 \\
          (8, 8, 9, 10, 10, -13, -18, -18, -27, 31) & 10 & 10 &   0 & -18 &      1 &  \boldsymbol{18} &   4 &   1 &   2 &   1 \\
             (1, 1, 1, -2, -2, -5, -5, 6, 6, 7, -8) & 11 &  8 &   1 &  -2 &      1 &                1 &   3 &   0 &   2 &   0 \\
            (1, -2, -2, -2, -3, 4, 4, -5, 6, 7, -8) & 11 &  8 &  -2 &   4 &      1 &               -2 &   3 &   1 &   1 &   1 \\
             (1, 1, 2, 2, 2, -4, -4, 7, -8, -9, 10) & 11 &  8 &   2 &  -4 &      1 &                2 &   2 &   2 &   1 &   2 \\
           (2, 2, 2, -4, -4, -5, 7, -8, 9, 10, -11) & 11 &  8 &   2 &  -4 &      1 &                2 &   3 &   0 &   1 &   0 \\
          (1, -2, -3, -3, -3, 5, 5, -7, 8, 10, -11) & 11 &  8 &  -3 &   5 &      2 &               -1 &   3 &   0 &   1 &   0 \\
            (3, 3, 3, -4, -4, 7, 7, -8, -9, -9, 11) & 11 &  8 &   3 &  -9 &      2 &                3 &   3 &   0 &   2 &   0 \\
          (1, 3, 5, -6, -6, -6, 8, -9, 12, 12, -14) & 11 &  8 &  -6 &  12 &      1 &               -6 &   3 &   1 &   1 &   1 \\
          (1, -2, 6, 6, 6, -7, 8, -9, -12, -12, 15) & 11 &  8 &   6 & -12 &      1 &   \boldsymbol{6} &   3 &   0 &   1 &   0 \\
          (1, 3, 3, 6, 6, 6, -7, -10, -12, -12, 16) & 11 &  8 &   6 & -12 &      1 &   \boldsymbol{6} &   2 &   2 &   1 &   2 \\
         (1, -2, -2, -2, 3, 3, 4, 4, -5, -5, -5, 6) & 12 &  9 &  -5 &  -2 &      1 &   \boldsymbol{7} &   3 &   0 &   2 &   0 \\
         (1, 1, -3, 4, 5, 5, 5, -6, -7, -7, -8, 10) & 12 &  9 &   5 &  -7 &      1 &                2 &   3 &   2 &   1 &   2 \\
         (1, 1, 1, -2, 4, -7, -7, -7, 8, 9, 9, -10) & 12 &  9 &  -7 &   9 &      1 &               -2 &   2 &   3 &   1 &   3 \\
        (1, 1, -3, -3, -5, -5, -5, 7, 7, 7, 9, -11) & 12 &  9 &  -5 &   7 &      1 &               -2 &   3 &   2 &   2 &   2 \\
     (1, -3, -3, -3, 4, 6, 7, 9, -10, -10, -10, 12) & 12 &  9 &  -3 & -10 &      1 &               13 &   3 &   0 &   1 &   0 \\
       (1, 1, 1, 3, 3, -5, 7, 7, -11, -11, -11, 15) & 12 &  9 &   1 & -11 &      1 &               10 &   3 &   1 &   2 &   1 \\
         (1, 1, 1, 3, 5, 5, -5, 5, -9, -9, -13, 15) & 12 &  9 &   5 &  -9 &      2 &                2 &   2 &   3 &   1 &   3 \\
  (1, -2, -2, 3, 6, -10, -10, -10, 13, 14, 14, -17) & 12 &  9 & -10 &  14 &      1 &               -4 &   4 &   2 &   2 &   2 \\
(1, -3, 9, -11, -13, -13, -13, 15, 15, 15, 21, -23) & 12 &  9 & -13 &  15 &      1 &               -2 &   3 &   1 &   1 &   1 \\
\bottomrule  
\end{tabular}
\end{scriptsize}
\caption{ Set of charges satisfying the Diophantine equations~\eqref{eq:NN3} together with the conditions enumerated in the text, for solutions with $N_{\textbf{chiral}}$ massive fermions, including 
the Dirac neutrinos masses allowed by the effective operator of D-$4+\delta$,
 the solution with unconditional stability through $\mathbb{Z}_{|s|}$ is highlighted with a bold font for the charge of $|s|=\boldsymbol{6},\boldsymbol{14}$. $N_D$ and $N_M$ are the number of massive Dirac and Majorana singlet fermions in each solution, while $G_D$ and $G_M$ are the corresponding number of maximum generations in the set.
}
\label{tab:sltns}
\end{table}

There, the list of $N$ integers in column `Solution' satisfy the Diophantine  equations~\eqref{eq:NN3} for an active symmetry if $m\ne0$ or for     a dark symmetry if $m=0$ (in such a case the number of new singlet chiral fermions is increased by 3). In each solution there are $N_{\text{chiral}}$ singlet chiral fermions, from which $N_\nu$ can develop effective Dirac neutrino masses from the operator in eq.~\eqref{eq:nmo56} with D-$(4+\delta)$. The remaining singlet chiral field acquires Dirac or Majorana masses from the Yukawa couplings with the complex singlet scalar, $S$. Finally,
 the solutions with unconditional stability are highlighted with a bold font in the column $s$ of Table~\ref{tab:sltns}. The last four columns correspond to the number of massive Dirac (Majorana) fermions $N_D$ ($N_M$) and the maximum number of generations for the Dirac (Majorana) fermions $G_D$ ($G_N$) in each solution.

 There are not types of solutions with a single massive Dirac or Majorana singlet fermion in several generations: $N_{DM}=1$ and $G_{DM}>1$, which would point out to a pure scotogenic realization of the Dirac neutrino mass operator without extra fermionic DM candidates. In fact, in general, most of the solutions feature multi-component dark matter with the lighter of all them protected by at least one remnant discrete symmetry, $\mathbb{Z}_{|s|}$. Each DM candidate can be the lighter in a sector with several generations, and for some cases highlighted with a bold $s$,  with at least two of them protected by  $\mathbb{Z}_p\otimes\mathbb{Z}_q$ with $p$ and $q$ coprimes and $|s|=pq$ .

\begin{figure}
    \centering
    \includegraphics[scale=0.7]{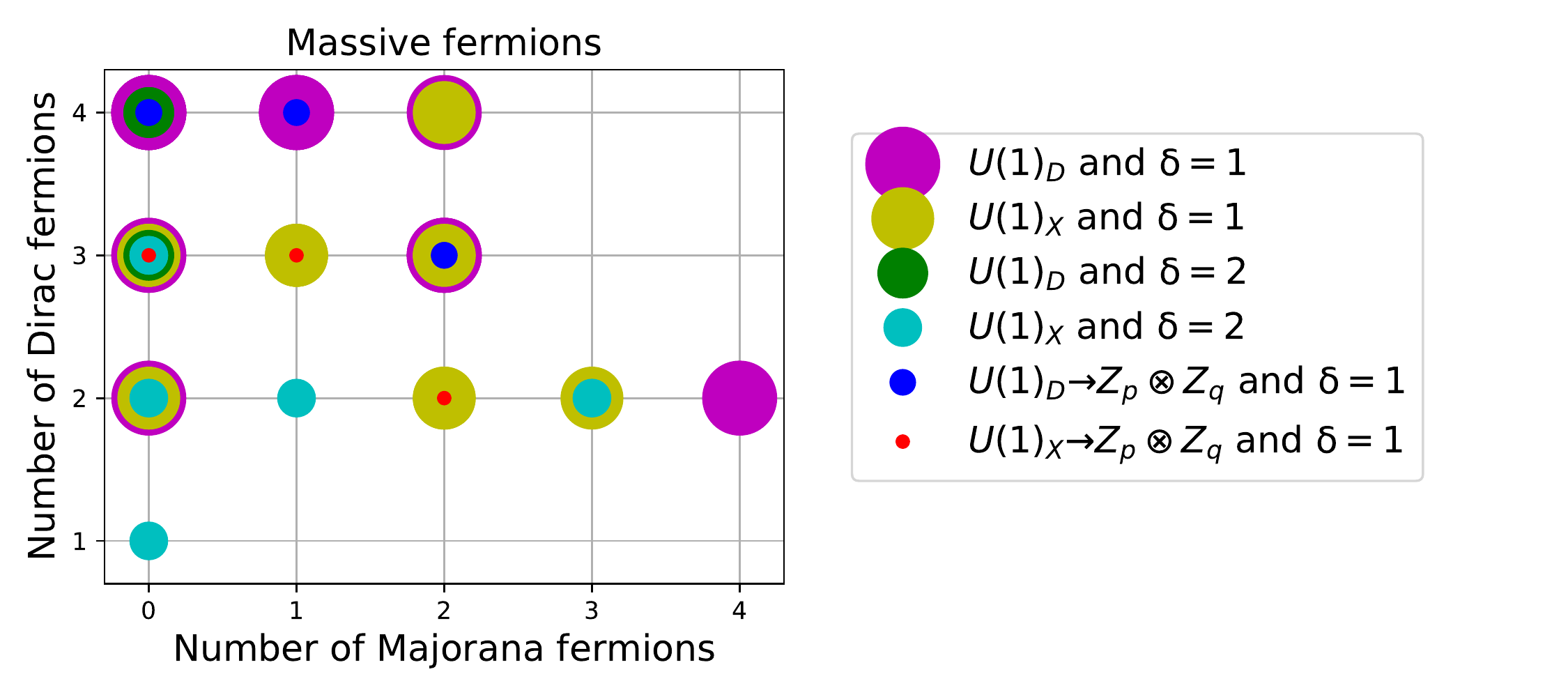}
     \caption{Number of massive Dirac and Majorana fermions of the full set of solutions in Fig.~\ref{fig:solutions}. The discs magenta and green (yellow and cyan) are for dark (active) symmetries which allows the effective Dirac neutrino mass operator for at least two right-handed neutrinos for D-$(4+\delta)$, with $\delta=1,2$ respectively. Similarly, the blue and red  points are the type of solution which satisfy the unconditional stability conditions for at least two DM candidates. }
    \label{fig:number}
\end{figure}

In Fig.~\ref{fig:number} it is illustrated the variety in which the number of Majorana and Dirac massive singlet fermions appears in the several types of solutions. We can have types of solutions with until four Majorana and two Dirac fermions for dark Abelian gauge symmetries of D-5 (magenta discs) and four Dirac fermions and two Majorana ones for dark or active Abelian gauge symmetries of D-5 (magenta and yellow discs respectively).

\begin{figure}
    \centering
    \includegraphics[scale=0.7]{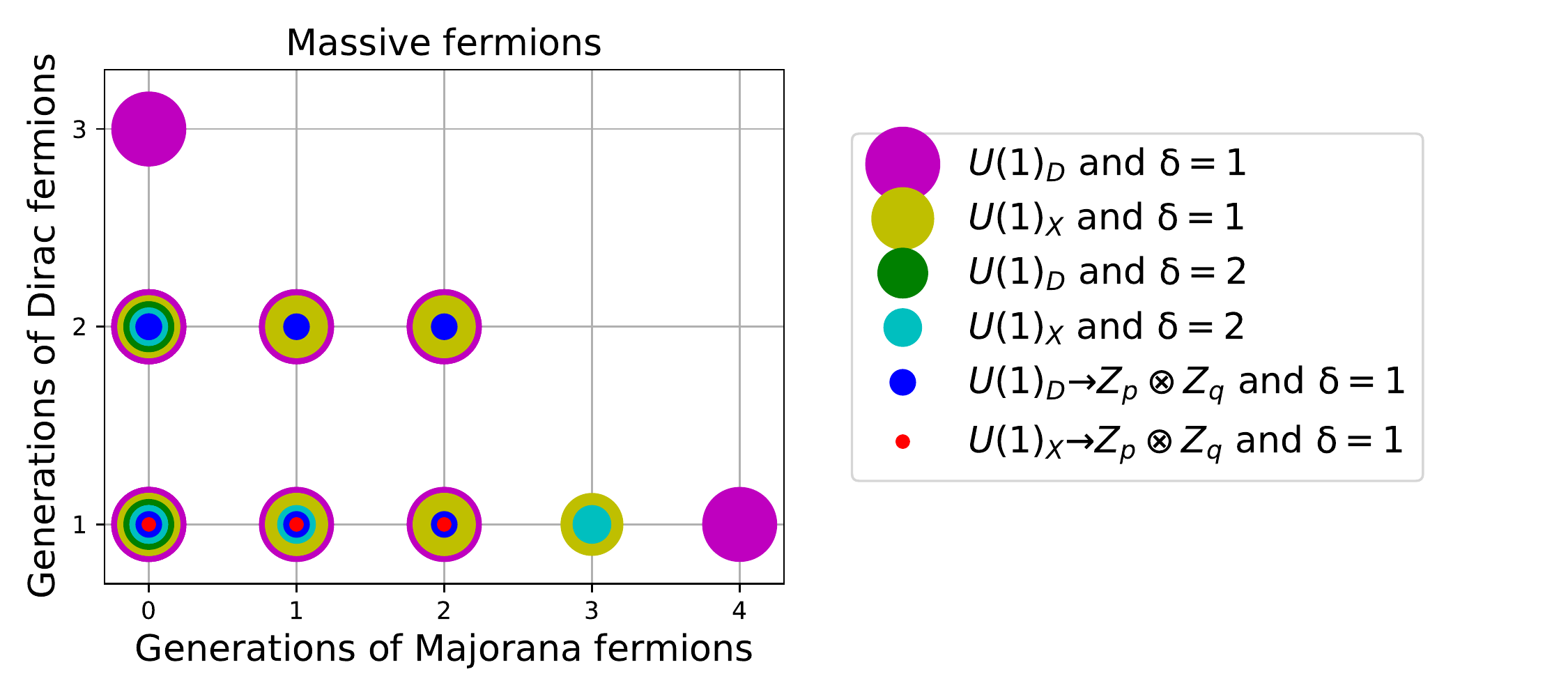}
    \caption{Same as Fig.~\ref{fig:number} 
    but for number generations of massive Dirac and Majorana fermions of the full set of solutions in Fig.~\ref{fig:solutions}. 
    }
    \label{fig:generation}
\end{figure}

Each massive singlet fermion can belong to a set of generations with the same quantum numbers but different masses. In Fig.~\ref{fig:generation} it is shown the way in which each dark matter candidate is grouped in generations for the several types of solutions. From a phenomenological point of view, this is important because this allows to have the rank of the effective neutrino mass matrix sufficiently high in scotogenic realizations of Dirac neutrino mass operator~\cite{Bernal:2021ezl}. Moreover, the interplay between several generations can boost the direct and indirect signals of the DM candidate~\cite{Herms:2021fql}. We can have types of solutions with until four generations of Majorana singlet fermions or with until three generations of Dirac singlet fermions, for dark symmetries of D-5 (magenta discs in the plot).

Note also that dark symmetries with $\delta=1$ and unconditional stability, represented as blue dots in figs.~\ref{fig:number} and~\ref{fig:generation}, are typically associated to a high number of massive singlet Dirac fermions, fig.~\ref{fig:number}, in sectors of at most two generations of massive Dirac or Majorana singlet fermions, fig.~\ref{fig:generation}.
On the other hand,  active symmetries with $\delta=1$ and unconditional stability, represented as red dots in figs.~\ref{fig:number} and~\ref{fig:generation},  are associated to at least two Dirac singlet fermions and less than three Majorana singlet fermions, fig.~\ref{fig:number}, without extra generations, fig.~\ref{fig:generation}.

In any scotogenic realization of the effective Dirac neutrino mass operator, a set of the heavy singlet fermions must participate directly in the generation of the light neutrino masses. In this way, the corresponding   fermion Yukawa couplings are also constrained by neutrino physics. Extra conditions are required to check if a solution can have a scotogenic realization.
In fact, all the solutions until $N=9$ for the one-loop scotogenic realization of the effective Dirac neutrino mass operator at D-5 and D-6 were presented in~\cite{Bernal:2021ezl}. As an example, the solution in the first row of Table~\ref{tab:sltns} was analyzed there, corresponding to one scotogenic realization of the D-5 operator at one-loop through a dark symmetry with two  independent Dirac fermionic dark matter candidates (magenta disc with coordinates $(0,2)$ in Figure~\ref{fig:number}). In general, for each type of solution we can have extra DM sectors which may be independent of the neutrino physics. 
It is worth noticing that in the absence of scotogenic conditions, the realization of the effective  Dirac neutrino mass operator is always possible for SM extensions with only extra scalars at tree-level~\cite{Ma:2021szi,Bernal:2021ppq}, one-loop~\cite{Saad:2019bqf,Calle:2021tez} and two and three-loops~\cite{Saad:2019bqf}.
Therefore, it is well motivated to make a systematic study of multi-component Dirac or Majorana fermionic DM, independent of the specific realization of the effective Dirac neutrino mass operator.
Moreover, since the neutrino physics is in general independent of the DM phenomenology, we expect that the General Neutrino Interactions (GNI) can be large and well suited to be explored in neutrino experiments~\cite{Calle:2021tez}.
Additionally, the dark symmetry models contains a dark photon which can be searched in both neutrino and direct detection experiments. Alongside, the active symmetry contains a $Z'$ with both quark and lepton couplings which have strong dilepton signals at the LHC.
It turns out that the $Z'$ portal also allows us to probe
the models with $m\ne0$ through modifications of the cosmological
history of the Universe, namely, via additional contributions from the right handed neutrinos to the effective number of relativistic degrees of freedom $N_{\text{eff}}$, with constraints that are competitive with the ones obtained from colliders~\cite{Calle:2019mxn}. 

In general, we can see that multi-component and multi-generational DM candidates are the trend for gauge Abelian extensions of the SM with massive singlet chiral fermions compatible with the effective Dirac neutrino mass operator of D-$(4+\delta)$, for $\delta>0$.

\section{Simplest realization: Dirac Zee model with Dirac fermionic DM}
\label{sec:zee}
As an illustration, consider  the type of solution as the one  in the second row of Table~\ref{tab:sltns}: $(3\c 3\c 3\c -5\c -5\c -7\c 8)$. This features the one with the least $N_{\text{chiral}}=4$, as summarized in Table~\ref{tab:pickedsltn}.
This solution is the simplest of  the types of solutions with $\operatorname{U}(1)_X$ and D-6 ($\delta=2$), represented by the cyan dot with coordinates $(0,1)$  in Figures~\ref{fig:number} and \ref{fig:generation}.

\begin{figure}
    \centering
    \includegraphics[scale=0.7]{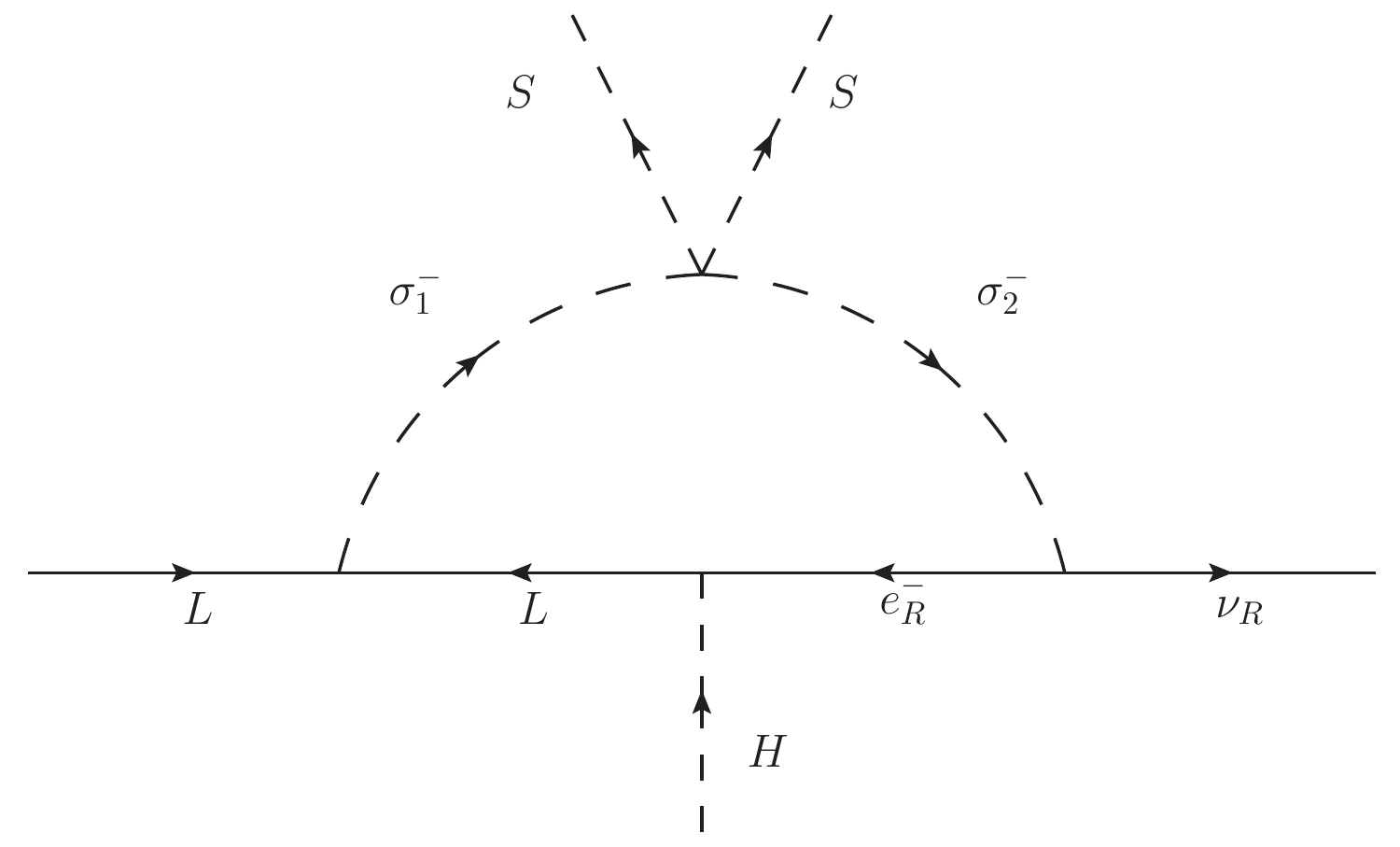}
    \caption{Diagram with the $X$-charge flux of the fields in the one-loop Dirac Zee model which realizes the effective Dirac neutrino mass operator at D-6}
    \label{fig:zee}
\end{figure}

The solution corresponds to an extension of the SM with $m=-3$ that for $h=0$ give rise to a local  $\operatorname{U}(1)_{B-L}$ until a global factor of $-1/3$: $u=d=-Q=1$ and $e=-L=-3$. This factor has been included in the column $\operatorname{U}(1)_{B-L}$ of Table~\ref{tab:pickedsltn}.  The effective Dirac neutrino mass operator of D-6 is compatible with two light Dirac neutrino masses through the right-handed neutrino, $\nu_R$, with a $(B-L)$-charge of $\nu=-5/3$. 
Without further conditions, this can be realized if we include two weak-singlet charged scalars, denoted as $\sigma^-_1$ and $\sigma^-_2$ in Table~\ref{tab:pickedsltn}. Then, we can have 
the one-loop realization of the effective Dirac neutrino mass operator at D-6~\cite{Saad:2019bqf,Calle:2021tez} illustrated in Figure~\ref{fig:zee}, which features the Dirac Zee model. From the diagram there, we have $\sigma_1=-2L$ and $\sigma_2=e+\nu=-2-2L$, with the corresponding $\operatorname{U}(1)_{B-L}$ charges in Table~\ref{tab:pickedsltn}. The neutrino related phenomenology of the model has been analysed in detail in~\cite{Calle:2021tez}, featuring large GNI along with charged lepton flavour violating processes, and strong dilepton signals at the LHC.

\begin{table}
  \centering
  \begin{tabular}{l|c|c|c|c                                  }\hline
    Field            &$SU(2)_L$     & $\operatorname{U}(1)_Y$ &$\operatorname{U}(1)_X$ &$ \operatorname{U}(1)_{B-L}$ \\ \hline
    $Q_i$            & $\mathbf{2}$ & $1/6$   & $L/3$ & $1/3$     \\
    $u_{Ri}$         & $\mathbf{1}$ & $2/3$   & $4L/3-3$ & $1/3$     \\
$d_{Ri}$         & $\mathbf{1}$ & $-1/3$   & $3-2L/3$ & $1/3$     \\    
    $L_i$            & $\mathbf{2}$ & $-1/2$   & $-L$ & $-1$     \\ 
    $e_{Ri}$         & $\mathbf{1}$ & $-1$   & $3-2L$ & $-1$     \\ 
    $\nu_{R \alpha}$ & $\mathbf{1}$ & $0$      & $-5$ & $-5/3$     \\ 
    $\psi_1$         & $\mathbf{1}$ & $0$      & $-7 $ & $-7/3 $  \\ 
    $\psi_2$         & $\mathbf{1}$ & $0$      & $8$ & $8/3$    \\
    $H$              & $\mathbf{2}$ & $1/2$    &  $L-3 $ &$0$  \\        
    $S$              & $\mathbf{1}$ & $0$      &  $1$ & $1/3$   \\\hline 
    $\sigma_1^-$     & $\mathbf{1}$ & $-1$    &  $2L $ & $2$  \\ 
    $\sigma_2^-$     & $\mathbf{1}$ & $-1$    &  $(-2-2L) $ & $-8/3$  \\\hline   
  \end{tabular}
  \caption{$X$ and proper $B-L$ normalized charges for the first solution in Table~\ref{tab:sltns}, $(3\c 3\c 3\c -5\c -5\c -7\c 8)$, for which $m=3$, $\nu=-5$, $\delta=2$ and therefore from eq. \eqref{eq:effcon}, $s=1$. For the column $\operatorname{U}(1)_{B-L}$ we fix $L=3$ and change $X\to X/3$. Moreover, $i=1,2,3$ and $\alpha=1,2$. The electroweak quantum numbers are also shown for each one of SM fields, the extra singlet chiral fermions, and the scalars in the last four rows.}
  \label{tab:pickedsltn}
\end{table}

In the new realization of the Dark Zee model presented here, we have the novelty that the model can also have a Dirac fermionic dark matter candidate. In fact,
the Lagrangian includes a new Yukawa term
 \begin{align}
     \mathcal{L}\supset & y_D \psi_1\psi_2 S^* +\text{H.c,}
 \end{align}
 with two singlet chiral fermions of $(B-L)$-charges $-7/3$ and $8/3$ respectively.  
 After the spontaneous symmetry breaking of the active gauge symmetry, this give rise to a Dirac heavy fermion mass
 \begin{align}
     \mathcal{L}_{\text{DM}}=M_\Psi \overline{\Psi}\Psi\,,
 \end{align}
 where $\Psi=\begin{pmatrix}\left(\psi_1\right)^\dagger & \psi_2\end{pmatrix}^{\operatorname{T}}$ and $M_\Psi=y_D \langle S\rangle/\sqrt{2}$.

Therefore, in contrary to the scotogenic models, the DM phenomenology can be studied independently of the neutrino physics. And therefore, previous analysis of simplified dark matter models with Dirac fermionic dark matter in extensions of the SM with $\operatorname{U}(1)_{B-L}$, fully apply here.
In particular, the annihilation cross section for $\overline{\psi}\psi\to {Z'}^* \to \overline{f}f$ is given in~\cite{Duerr:2015wfa}. As we implemented this model in SARAH~\cite{Staub:2013tta}, we can have the relic density calculation directly from micrOMEGAs~\cite{Belanger:2018ccd}.

The corresponding spin-independent direct detection cross section of dark matter per nucleon is given by~\cite{Duerr:2014wra,Duerr:2015wfa}
\begin{align}
    \sigma_{\psi N}
    =&\frac{\mu_{N}^{2}}{\pi}\frac{g_X^4}{M^4_{Z'}}\left(\psi_1-\psi_2\right)^2L^2 \,, 
\end{align}
where $\psi_1=-7/3$, $\psi_2=8/3$ and $L=1$, are the $\operatorname{U}(1)_{B-L}$ charges in Table~\ref{tab:pickedsltn}, and
\begin{align}
    \mu_N=\frac{M_N M_\Psi}{M_N+M_\Psi}\,,
\end{align}
is the reduced mass.

The cross section can be written as
\begin{align}
    \sigma_{\psi N}^{\mathrm{SI}}\left(\mathrm{cm}^{2}\right)=4.8 \times 10^{-47}\left(\frac{\mu_N}{1 \mathrm{GeV}}\right)^{2}\left(\frac{0.1}{g_X}\right)^{4}\left(\frac{M_{Z'}}{4\ \text{TeV}}\right)^{4} \left(\psi_1-\psi_2\right)^2L^{2} \mathrm{~cm}^{2}\,. 
\end{align}

\begin{figure}
    \centering
    \includegraphics[scale=0.6]{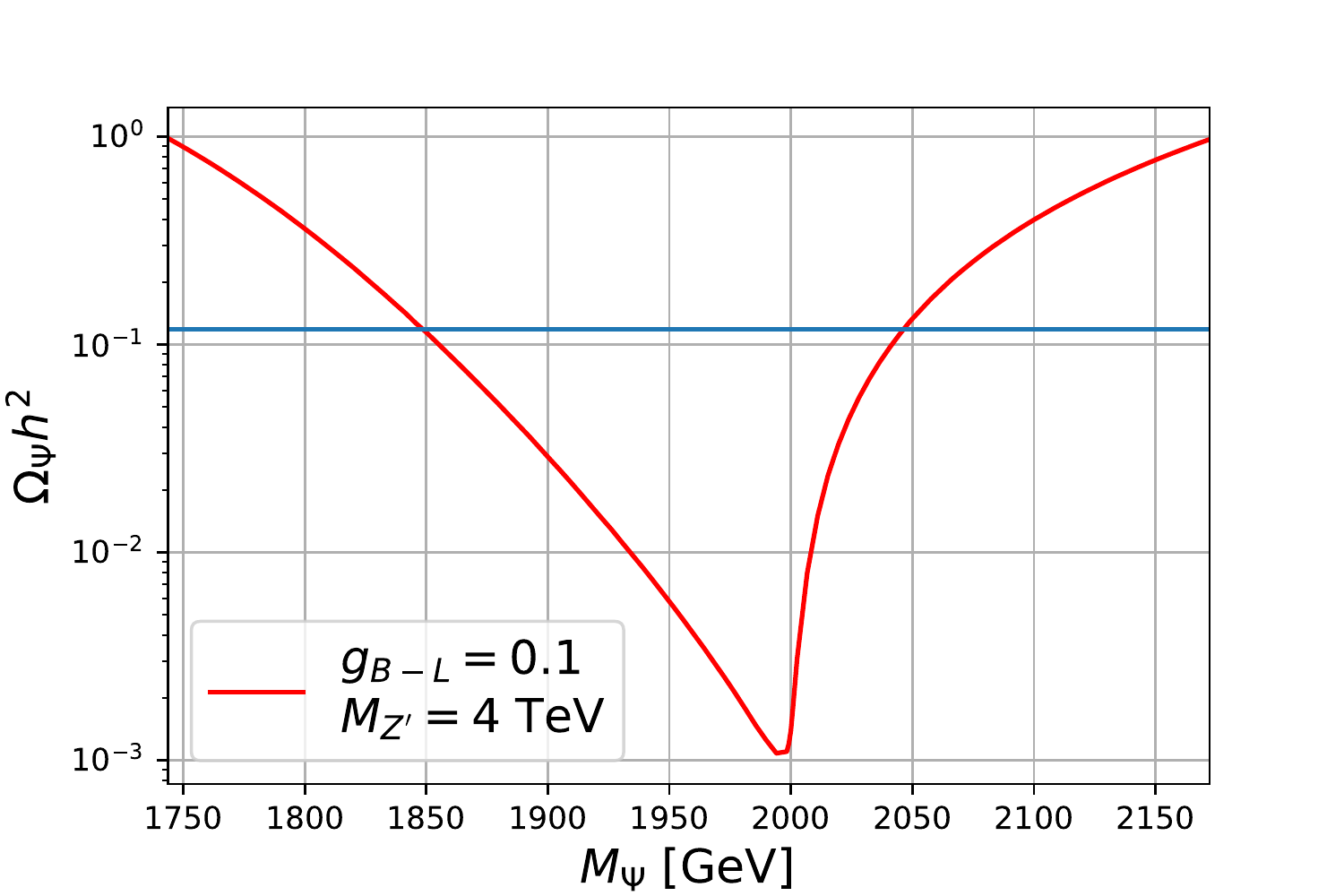}
    \caption{Dirac fermionic Dark matter relic density as a function of its mass for $\operatorname{U}(1)_{B-L}$ with copuling $g_{B-L}=g_X=0.1$, $M_{Z'}=4\ \text{TeV}$, $m_S=5\ \text{TeV}$. }
    \label{fig:dm}
\end{figure}

We can illustrate the DM phenomenology for a specific benchmark point  which is compatible with collider, $\Delta N_{\text{eff}}$~\cite{Calle:2019mxn} and the current direct detection currents constraints~\cite{PandaX-4T:2021bab}. We choose $g_X=0.1$, $M_{Z'}=4\ \text{TeV}$, $m_S=5\ \text{TeV}$. 
For this specific benchmark point, the main constraint arises from the current upper bound on $\Delta N_{\text{eff}}$ for which $M_{Z'}\gtrsim 40\ \text{TeV}$. This kind of constraint has the potential to probe the full model with the next generation of CMB experiments~\cite{Calle:2019mxn}. 
From Figure~\ref{fig:dm} we can see that the proper relic density can be obtained for $M_\Psi\approx 1840 \text{ GeV}$ or  $M_\Psi\approx 2050 \text{ GeV}$

\section{Conclusions} 
\label{sec:con}
We found around one thousand solutions to the anomaly free conditions of local Abelian extensions of the SM with $N$ 
integers and $N_{\text{chiral}}$  right handed singlet fermions which include the ones forming the effective set of Dirac neutrinos and a dark sector with massive fermions. We classify the solutions in 48 types in Table~\ref{tab:sltns} depending if the symmetry is dark ($N_{\text{chiral}}=N$) or active ($N_{\text{chiral}}=N-3$), the dimension of the effective Dirac neutrino mass operator ($4+\delta$), the number of independent dark matter candidates, and the number of generations of each massive fermion. The scalar realizations of the effective Dirac neutrino mass operator feature a set of parameters which explain independently the neutrino oscillations and the phenomenology of a multi-component and multi-generational dark matter sector. This allows for large GNI and charged lepton flavor violation observables while keeping the features of the simplified dark matter models. We have illustrated this separation explicitly with a simple one-loop Dirac Zee model which has a Dirac fermionic dark matter candidate. We could expect enhanced effects in scalar realizations of the effective Dirac neutrino mass operator at two-loops and three-loops~\cite{Saad:2019bqf}, which are not yet fully explored in the literature.

\section*{Acknowledgments}
The  work  of  DR  is  supported  by  Sostenibilidad  UdeA,  and the UdeA/CODI  Grants 2017-16286 and 2020-33177.

\bibliographystyle{apsrev4-1long}
\bibliography{biblio}

\end{document}